\documentclass[final,5p,times,twocolumn,authoryear]{elsarticle}

\usepackage{amssymb}

\usepackage{lineno}
\usepackage{hyperref}

\journal{Astronomy and Computing}

\begin{document}

\begin{frontmatter}



\title{Toyz: A Framework for Scientific Analysis of Large Datasets and Astronomical Images}


\author{Fred Moolekamp and Eric Mamajek}

\address{Department of Physics \& Astronomy, University of Rochester, Rochester, NY, 14627-0171, USA}

\begin{abstract}
As the size of images and data products derived from astronomical
data continues to increase, new tools are needed to visualize and
interact with that data in a meaningful way. 
Motivated by our own astronomical images taken with the Dark Energy Camera (DECam) we
present \emph{Toyz}, an open source Python package for viewing and analyzing images
and data stored on a remote server or cluster. Users connect to the \emph{Toyz} web
application via a web browser, making it an convenient tool for students
to visualize and interact with astronomical data without having to install any software on their
local machines. In addition it provides researchers with an easy-to-use tool that allows them
to browse the files on a server and quickly view very large images ($>$ 2 Gb) taken with
DECam and other cameras with a large FOV and create their own visualization tools that
can be added on as extensions to the default Toyz framework.

\end{abstract}


\end{frontmatter}

\section{Introduction}

In the past, large scientific datasets were used mainly by large collaborations
while independent researchers worked with much more manageable volumes
of data. 
Over the past few years we've been entering a new paradigm
where very large sets of data are available to (and at times even generated by)
much smaller groups. This abundance of data has highlighted a shortage
of scientific tools to store, organize, analyze, and visualize that
data. Fortunately this problem overlaps with the needs of
the industrial community at large and in the past decade there has
been a lot of work by traditional scientists, data scientists, and
software engineers to develop software to aid researchers in dealing
with this new (and rewarding) problem. 

Unfortunately much of the current work in astronomy is often on
the fringe of what is possible and has been done before, meaning
the types of data we work with poses new challenges, which in
turn create a need for new tools 
\citep{Merenyi2014,Gopu2014,Lins2013,Loebman2014,Federl2012,Federl2011}. 
Ideally these new tools should be built on existing
frameworks that are under active development by software engineers
to minimize the effort from research scientists while taking advantage
of the latest technologies and updates to existing codes. The Python
language has become a fertile ground for rapid software development
and with the creation of a vast array of modules for scientific
image and data processing like \emph{numpy} \citep{Walt2011}, \emph{scipy} \citep{Jones2015}, 
\emph{pandas} \citep{Mckinney2010} and \emph{scikit-image} \citep{vanderWalt2014}; 
machine learning modules like \emph{scikit-learn} \citep{Pedregosa2011};
statistics and modeling packages like \emph{scikits-statsmodels}, 
\emph{pymc} and \emph{emcee} \citep{Foreman2013}, and what has become the de facto astronomy
python project \emph{astropy} \citep{Robataille2013} and its affiliated packages.

While many of the tools listed above are useful for astronomers, data scientists,
and software engineers; there is a great divergence when it comes to tools for
visualization. Much of the interactivity and visualization work done in the realm of data science
and software development tends to be focused on web frameworks like 
\emph{jQuery Ui}, \emph{Highcharts}, \emph{D3.js}  and even more advanced
libraries using webGL like \emph{PhiloGL}, \emph{pathGL} and many others; or 
R libraries like \emph{ggplot2} \citep{Hadley2009}.
Contrast this with astronomy where programs like
\emph{ds9} \citep{Joye2003} that are used primarily by astronomers with few updates
and changes over the past decade. Several recent python packages have been 
created to help bridge the gap between professional visualization tools and those
available in astronomy: \emph{GLUE} \citep{Beaumont2014} provides a rich GUI for interacting
with data sets and images and \emph{Ginga} is one of the most advanced frameworks
for viewing and interacting with FITS images.

The disadvantage of using any of the visualization tools in astronomy 
mentioned above is that to run efficiently all of them must be run on a local machine with data
stored locally. With new instruments like the Dark Energy Camera (DECam) that create
2Gb images (~.5Gb compressed) and over 1Tb of data products per night
\citep{Valdes2014, Flaugher2012}, it's no longer feasible to 
store an entire observing run (or even a single night) on a laptop or PC. Recognizing
the need for a server side image viewer several groups have been independently 
developing web applications to serve images from a remote server to a client with only
a web browser installed including \emph{VisiOmatic} \citep{Bertin2015}, 
\emph{Data Labs} \citep{Fitzpatrick2014}, 
and now \emph{Toyz}. \emph{VisiOmatic} is an open source 
web application running on an Apache web server
with an IIPImage \citep{Pillay2014} server to display large images in a browser using a so
called ``slippy map'' implementation (similar to Google maps). In addition to 
viewing images, the \emph{VisiOmatic} client also enables users to interact with the 
image including marking point sources and plotting slices of the data. 

When we first began to analyze our own DECam images, which took up over 1 Tb of
disk space on our server, we realized that in order to view the images and analyze 
the catalogs we created with them that we would need a new tool, preferably one 
that could run on the server
storing the data and allow a platform independent way for users to connect to the data and
interact with it. This was our initial motivation for creating a new python package 
called \emph{Toyz}, 
which seeks to combine the best of all of the software discussed so far: the 
remote image viewing of \emph{VisiOmatic}, the interactivity of \emph{GLUE} and \emph{Ginga}, the 
astronomical tools of \emph{astropy}, and the convenience of doing it all in a single
framework built on existing Python and HTML5 software maintained by 
computer scientists. Because
\emph{Toyz} is a framework, not an application, it is designed to be easily customized
by end users for their specific scientific needs but easy enough to use that
a class of undergraduates could use it for analyzing their data without having to
install any software on their home computers. One of the guiding principles of
\emph{Toyz} is that an undergraduate student should be able to install \emph{Toyz} and begin
analyzing data on his/her first day!

In this paper we highlight the various functions of \emph{Toyz}.
Section \ref{Toyz} describes the core \emph{Toyz} package that allows users to 
view images and interactive plots in their browsers, section \ref{Affiliated} describes
the affiliated package \emph{Astro-Toyz} that incorporates astronomy specific
tools including WCS and interactive tools for the image viewer, and section \ref{Future}
describes future plans for integrating \emph{Toyz} with other software packages.

\section{Toyz}\label{Toyz}

\emph{Toyz} can be thought of as a platform-independent
tool for visualizing and interacting with large images or catalogs
of data. Instead of trying to create a one-size-fits-all application,
\emph{Toyz} is designed to be an open source framework that scientists 
can customize to fit their own research needs. 

\begin{figure}[h]
\includegraphics[width=\linewidth]{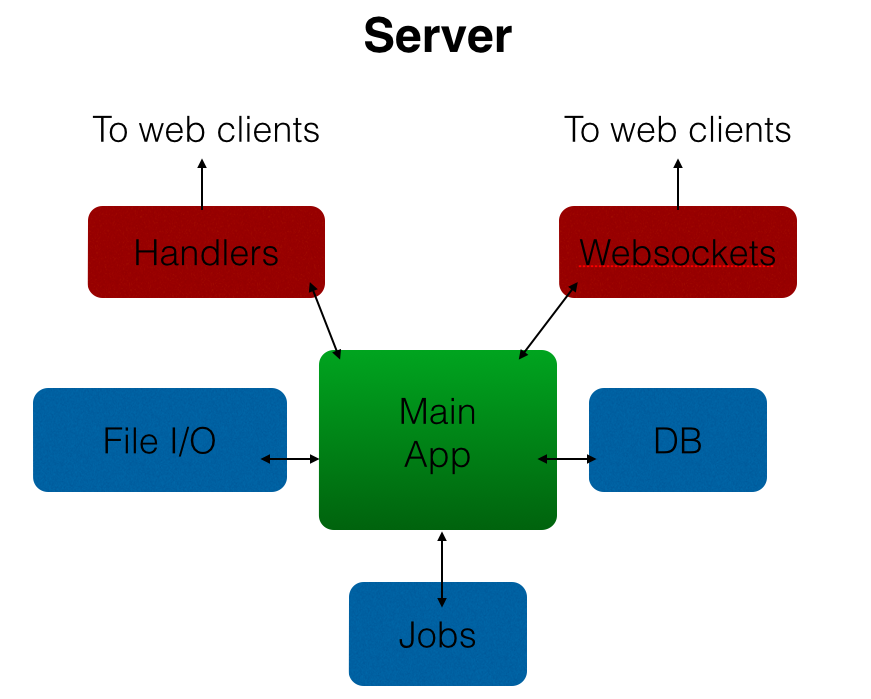}
\caption{Server Data Structure}
\label{fig:Server}
\end{figure}

A graphical representation of the server is shown in Figure \ref{fig:Server}.
The web application at the heart of \emph{Toyz} is built on the Tornado web
framework \citep{Darnell2015}, a python library originally written by FriendFeed
as the backend for their social media website. User authentication
is done via HTML handlers built into Tornado while most other
communications between the server and client are done via WebSockets \citep{hickson2011websocket}: 
a bi-directional protocol that uses an HTML handshake to setup
an open communication between the server and the client without the
need for constant polling by the client to get the status of a job.
Similar technology is used for a variety of websites and web applications
including Jupyter (formerly iPython) notebooks \citep{Ragan2014}. 
A separate module handling file I/O provides an API to load data from a 
variety of formats (see Section \ref{sub:Plots}). The file I/O module
is written to allow users to create affiliated packages or extensions 
that allow users to create custom classes for loading additional data types
not currently supported by \emph{Toyz} with minimal coding.

Each time a new connection is made to the server a new process called a \emph{session} is
spawned using python's multiprocessing module. All of the variables and
methods defined in a session will be stored until the user closes the browser and
disconnects from the server. All of the jobs sent from a client to the server are verified
for authenticity and put in a queue to be run for the correct session. Once a job is
completed, a response in the form of a JSON object is sent to the browser 
that at a minimum contains a status key (indicating whether or not the job 
completed successfully or encountered an error) and often additional 
keyword arguments generated by the function.

\begin{figure}[h]
\includegraphics[width=\linewidth]{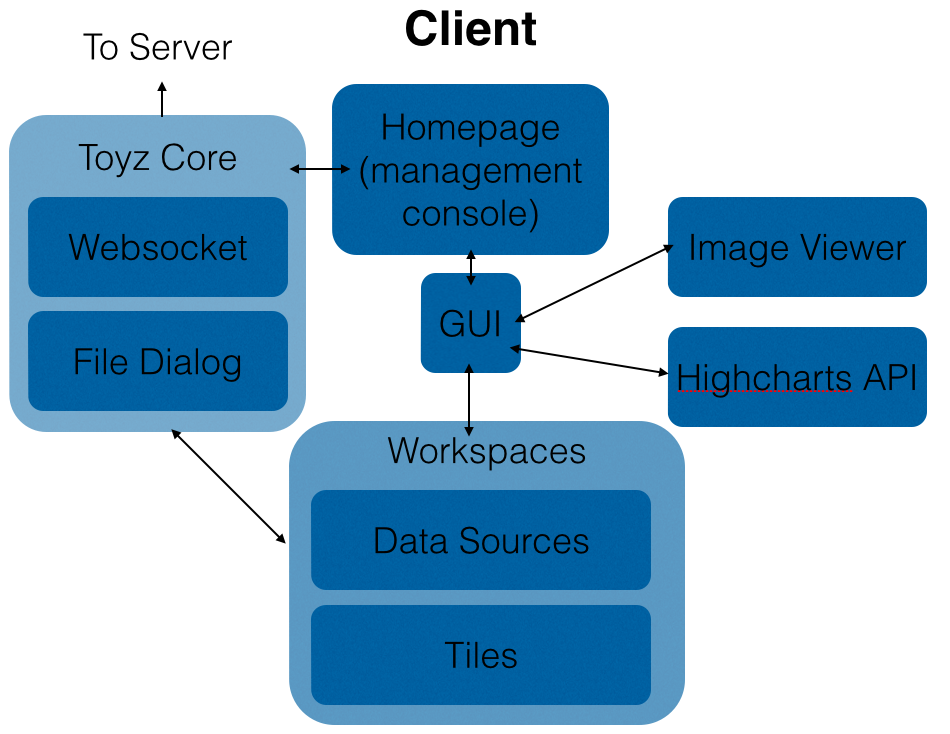}
\caption{Client Data Structure}
\label{fig:Client}
\end{figure}

\begin{figure*}
\includegraphics[width=\textwidth]{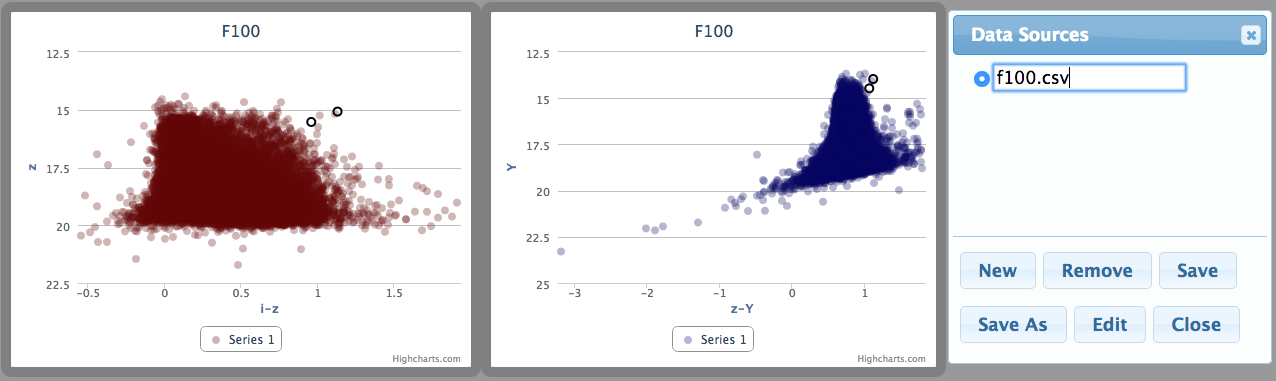}
\caption{Workspace with two different color-magnitude plots loaded from the same
    data source. When points are selected in one of the plots \emph{Toyz} will automatically
    select the same points in every additional plot generated from the same data 
    source on the server.}
\label{fig:plots}
\end{figure*}

On the client side all communications are pushed through a single function that
maintains information about the current session 
(a graphical representation of the client is shown in Figure \ref{fig:Client}). 
When initialized the user can
choose how errors and warnings that might occur while running a job are handled
as well as what actions to take when various types of responses are returned.
A file dialog is also initialized that allows users to browse the directory tree
on the server, functionality that is not incorporated into web browsers for obvious
security reasons.
The default homepage when a user logs onto the server is a management console that allows
one to set shortcut paths and allows administrators to set user and group permissions 
(see Section \ref{sub:Security}). 

To assist users in developing their own custom tools a GUI module parses JSON
objects (or python dictionaries) to build interactive tools like drop boxes, sliders,
and buttons without the need for javascript or css code. All of the menus and
controls in \emph{Toyz} have been generated using the same GUI framework, which  is 
thoroughly documented on the website at 
\url{http://fred3m.github.io/toyz}
along with several examples.

The remainder of this section discusses additional built-in features of 
\emph{Toyz}.

\subsection{Workspace Environment}\label{sub:Workspace}

The main working environment in \emph{Toyz} is referred to as a \emph{workspace}.
On the client side a workspace is simply a blank webpage that allows
users to add a collection of customizable tiles. Each tile is associated
with some functionality, such as displaying an image or plot, and
can be moved and resized in the browser window. On the server side
a workspace is an environment spawned as a new python process for
each window (or tab) in a user's browser. This environment is completely
separate from the web application (which handles connections to and
from the server) where the state of a user's variables are stored for
the duration of the connection. Because each workspace is a separate process,
\emph{Toyz} is able to take advantage of all of the processing cores on a server, meaning that
if the number of processors scales with the number of simultaneous users,
large classes and groups should be able to access the same \emph{Toyz} server with
no reduction in performance.

While \emph{Toyz} comes with two default tile types: \emph{Highcharts} plots 
(see Section \ref{sub:Plots})
and image viewers (see Section \ref{sub:Viewer}), a template is included with
the source code to allow users to create their own custom tiles with
access to all of the variables of the workspace on the server and the client. 
In addition, users are able to save the workspace by generating
a url that will load a saved workspace for the user and any collaborators
with whom the url and permissions are shared.

\begin{figure*}
\includegraphics[width=\textwidth]{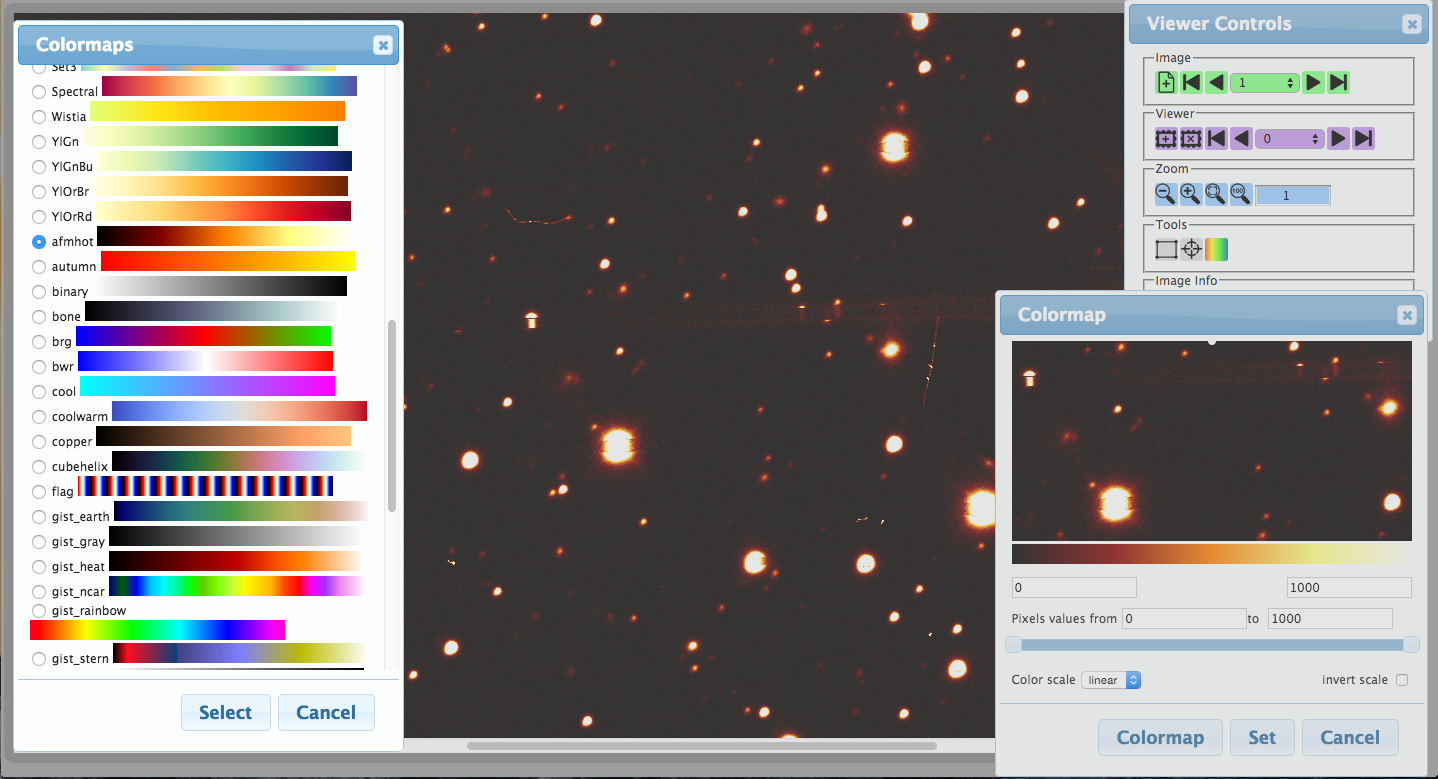}
\caption{\emph{Toyz} Image Viewer. For FITS files the full catalog of matplotlib colormaps is
    available as well as linear and log scaling and inverse color maps. To save upload time
    only a small tile in the center of the image is used to modify a colormap before it is
    applied to an entire image.}
\label{fig:viewer}
\end{figure*}

\subsection{Data Connectivity and Interactive Plots\label{sub:Plots}}

The gold standard open source package for data visualization in python is 
\emph{GLUE}, a python package which allows users to load a series of data
sets into memory and provides a GUI for plotting connected data sets
and images in tiles on the viewing window. The only current drawback
to \emph{GLUE} is that the data sources must be stored locally, which is
not always practical for the reasons mentioned earlier. While a future
version of \emph{Toyz} seeks to integrate with \emph{GLUE} and extend all of its
functionality to the browser, the current incarnation ports some of
the most important features, including the ability to create
linked plots.

Users are provided an interface to load a variety of data sources onto the server.
By default \emph{Toyz} will load any format that is integrated with \emph{pandas} including
SQL databases, HDF5, and text files as well as \emph{numpy} binary files and text files
that can be opened using standard python I/O functions.
Until the user tries to do something with the
data (like plot it in the browser) it remains solely on the server, saving time and
bandwidth. It is also possible to extend the available file types by adding a custom
module, for example the \emph{Astro-Toyz} package extends the available file types to
FITS tables, VOTables, and all of the other file formats that can be read from
\emph{astropy} Tables.

To interact with the data \emph{Toyz} provides a GUI to \emph{Highcharts}, 
an open source javascript 
library that allows users to display interactive plots in a web browser and is free
for academic and personal use (and reasonably priced for commercial use). \emph{Highcharts}
includes functionality to select data points as well as drill down, zoom into subsets of
a plot, and display information about a highlighted point. 

\emph{Toyz} includes an interface to choose columns from a data source loaded on the server
and create a plot using a subset of the \emph{Highcharts} API. The user can choose the title,
axis labels, tick marks, grids, line styles, marker colors and shapes, and various 
other features to make plots easier to view without any programming necessary. 
Each plot is created in a new
tile in the workspace and all of the plots connected to the same data source are linked
together so that selecting a point (or collection of points) in one of the plots will 
select the same point(s) in all of the other plots, making it easier to view high dimensional
or ``wide'' data (see Figure \ref{fig:plots}). Consistent with the \emph{Highcharts} API, 
multiple data sets can also be plotted
on the same chart. While \emph{Toyz} lacks some of the more advanced features of 
\emph{GLUE} at the 
moment, like merging data sets and linking table columns to image axes, it provides
a previously unavailable method to quickly explore data stored remotely.

The data source API is modular so that while \emph{Highcharts} is currently the only supported 
plotting library it would be straightforward to add an interface for other packages
like D3.js or webGL support.

\subsection{Image Viewer\label{sub:Viewer}}

The image viewer was initially developed to view collections of large
astronomical images that were too large to fit on a local hard drive.
Until recently the domain of viewing astronomical images rested on
software that had to be installed on a users machine and could only
efficiently view images stored locally. Even larger
detectors such as MOSAIC \citep{Pogge1998} at 8Kx8K px and the One Degree Imager (ODI) 
\citep{Jacoby2002} at ${12k \times 12k}$ px
produce images small enough that a single nights observations can
easily be stored on a notebook or PC and viewed on one of the existing
viewers. Newer cameras like the Dark Energy Camera (DECam) \citep{DePoy2008} with
a $30k \times 30k$ FOV enable a single observer (or team of scientists)
to generate over 1Tb of processed data in a single night, making 
all existing open source tools (other than \emph{VisiOmatic}) inconvenient
and inefficient.

Of course large FITS images are not the only types of images worth
viewing in a browser. One of the byproducts of data analysis is often
a large collection of plots that are generated wherever the data is
stored and processed (in this case on a server).
Depending on the software used, the filetype of these plots can vary
and it is also useful to have the ability to view these plots without
copying them from the server to the local machine. \emph{Toyz} uses Pillow, 
a fork of the Python Imaging Library, which allows users to
view a wide array of image formats including bmp, eps, jpeg, png,
and tiff files, as well as \emph{astropy} to load FITS images. Users are given
the option as to whether an image is loaded as a mosaic of tiles or
as a single image (which might be more useful in the case of
small images like plots).

The image viewer consists of standard tools such as scaling, panning and centering,
as well as a few additional handy features. Since many modern astronomical and
scientific images contain multiple frames, a set of controls allows users to easily
browse through the different frames of a single FITS image 
(see Figure \ref{fig:viewer}). It is also possible to
have multiple viewer frames loaded at the same time, making it easy to switch or 
``blink'' between different images (see Section \ref{sub:Astro-Toyz} for more on 
blinking and other tools specifically related to astronomical FITS images).

The viewer is also designed to be customizable in that end users can add their
own controls to the toolbar and even create their own custom image loaders. 
Since the viewer is also just a tile in a workspace, it is possible to have
multiple viewers loaded at the same time and in the future it should even be
possible to link the images to catalogs similar to \emph{GLUE}.

\subsection{Security\label{sub:Security}}

\begin{figure*}
  \includegraphics[width=\textwidth]{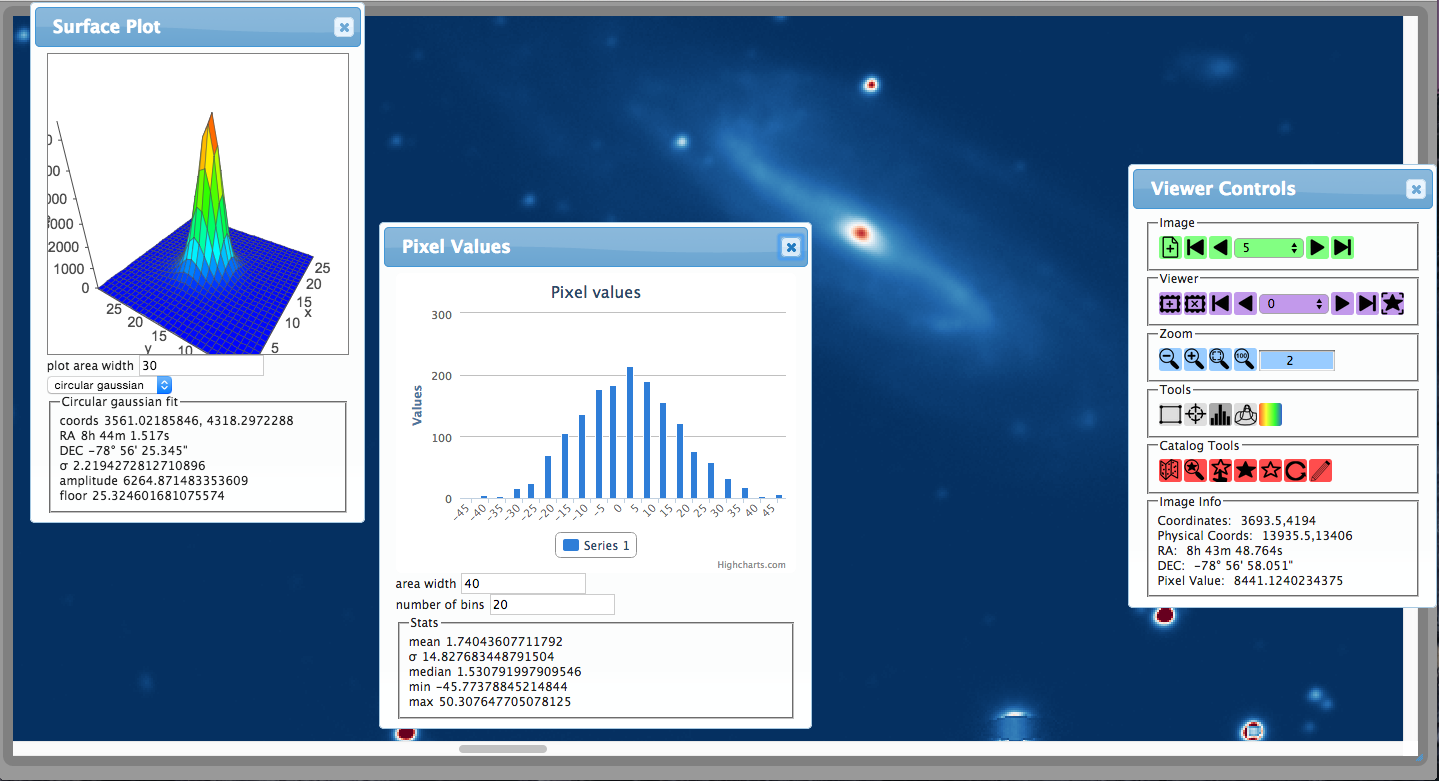}
  \caption{Image displayed using the \emph{Astro-Toyz} image viewer. DECam image in the vicinity of the spiral galaxy ESO 18-13.}
  \label{fig:astro-viewer}
\end{figure*}

The typical \emph{Toyz} install doesn't require much in the way of security.
The recommended install of \emph{Toyz} for research purposes is to install
the application on a server, log on to the server using a secure
shell and forward the port \emph{Toyz} is running on from the server to the
local machine. As long as the server is located behind a firewall
the need for security is limited, especially if there is no reason
for the users in the group to keep their analyses or data private
from one another. In this simplest use case each user can be added to
an admin group, allowing them access to all of the files on the
server (that the account running the instance of \emph{Toyz} has permission
to access) and run any python module that conforms to the \emph{Toyz} standard.

\begin{figure}[h]
\includegraphics[width=\linewidth]{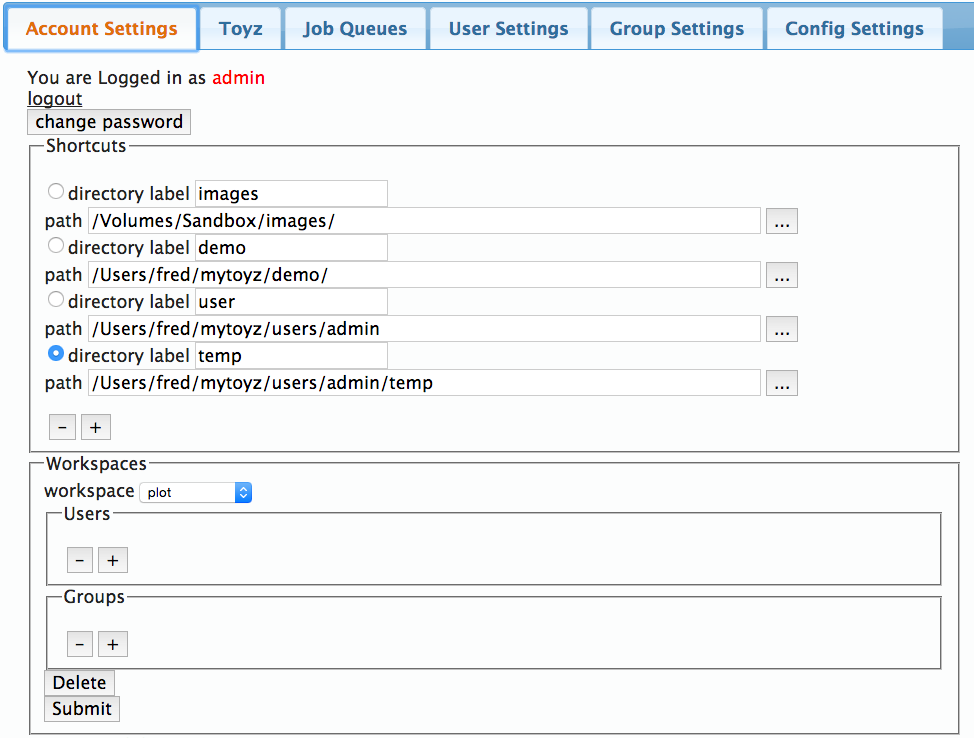}
\caption{Management Console}
\label{fig:console}
\end{figure}

In other scenarios, for instance groups working with confidential
data or classrooms where students shouldn't have access to each others
data or analyses, it is necessary for each user to have his/her own
account. \emph{Toyz} provides an admin console webpage that allows administrators
to create new users and groups as well as change their permissions for a
wide variety of features. By default, each user outside the admin
group doesn't have permission to view any directories outside of the
default directory created for him/her when his/her account is created
and only administrators can change those permissions. This allows
some directories to be shared by specified users or groups while remaining
private from others.

More importantly, because \emph{Toyz} acts as a GUI to the entire python
library, without specific precautions taken a user would be able to
run any python module, giving them the ability to run arbitrary code
on the server. To combat this \emph{Toyz} users and groups outside the admin
group can only run python modules they are specifically given permission
to run and no user is allowed to run a python module or function that
does not conform to a specific standard given in the \emph{Toyz} documentation.

\section{Affiliated Packages and Extensions}\label{Affiliated}

A template is included with the \emph{Toyz} source code that allows users
to create their own custom \emph{Toyz}. This allows them to create custom web pages,
new workspace tile types, as well as python data types, classes, and functions for any
purpose that their group sees fit, as long as they conform to the
standards specified in the template. Many of the built-in \emph{Toyz} functions
can also be wrapped, as in the example of the Astro-Viewer (see Section
\ref{sub:Astro-Toyz}), where even the control panel of the viewer
has been created in such a way as to allow users to add their own
buttons and controls. The \emph{Toyz} website has a section for affiliated
\emph{Toyz} called the Toy Box which will host links to packages created
by other users or groups built on the \emph{Toyz} framework.

\subsection{Astro-Toyz\label{sub:Astro-Toyz}}

To demonstrate the flexibility of \emph{Toyz} as well as support our own
research we developed an affiliated package called \emph{Astro-Toyz}. While
\emph{Toyz} was designed for general data visualization and analysis, \emph{Astro-Toyz}
is designed specifically for the analysis of astronomical data. It contains
add-ons to the image viewer that displays world coordinates,
plot histograms or surface plots (similar to imexam in IRAF \citep{Tody1986}),
and align images in separate viewer frames to the same coordinates and
scaling so that images can be blinked (see Figure \ref{fig:astro-viewer}).

In the long run the goal is for \emph{Astro-Toyz} to be a front end for \emph{astropy},
providing a GUI for users to make use of \emph{astropy} tools and affiliated
packages such as converting WCS coordinates, object detection, matching
to source catalogs from online sources like Vizier \citep{Ochsenbein2000} by using 
\emph{astroquery}, performing precision astrometry and photometry, 
and various other tasks supported in the \emph{astropy} universe. 
We hope to entice the large community of developers
who maintain other \emph{astropy} packages to add their own interfaces onto
\emph{Astro-Toyz} to broaden its scope. At that time \emph{Astro-Toyz} will be capable of
being implemented in undergraduate astronomy classes, allowing users
to perform all of their analysis from their own computers without
installing any software.

Currently \emph{Astro-Toyz} isn't quite ready for that level of interaction
but does provide additional tools to the \emph{Toyz} image viewer that give
users access to a number of advanced features including displaying
WCS and header info, access to the full matplotlib \citep{Hunter2007} catalog
of color maps, WCS alignment between images and blinking between multiple
aligned images for moving object detection.

\subsection{Extensions}\label{sub:Extensions}

\begin{figure}[h]
  \includegraphics[width=\linewidth]{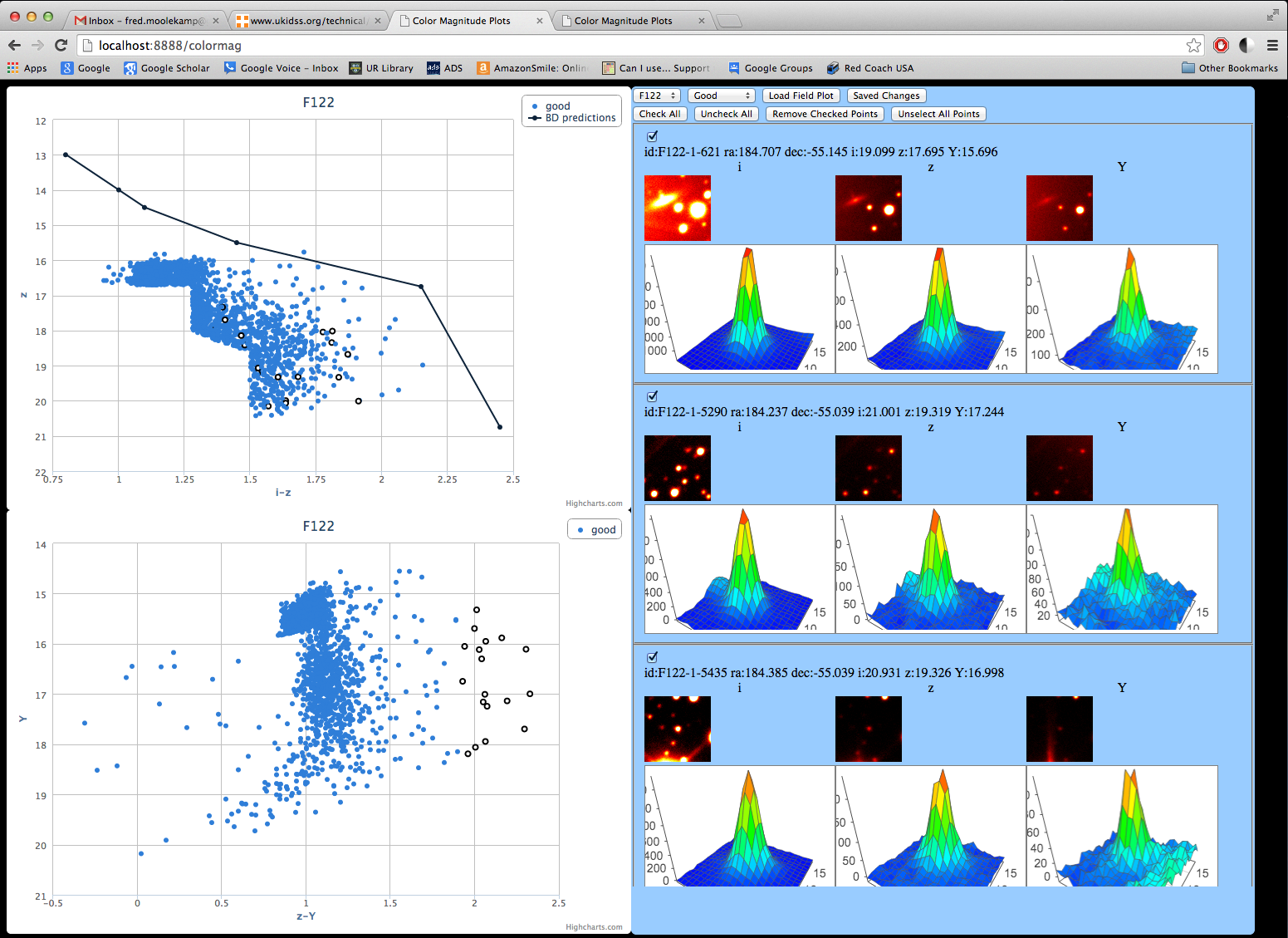}
  \caption{Custom webpage to interact with a point source catalog. 
    On the left are interactive \emph{Highcharts} plots. 
    On the right is a custom tile created to display an image and surface plot 
    from images taken of the same field with three different filters. Each set of 
    images on the right corresponds to a different source selected in the \emph{Highcharts}
    plots and can be removed if the detection is an artifact.}
  \label{fig:colormag}
\end{figure}

At times it may also be useful to write a short module to extend the functionality
of \emph{Toyz} (or a \emph{Toyz} affiliated package) for a specialized task. 
Figure \ref{fig:colormag} shows an example
of a custom workspace tile that loaded images and source information for 
point sources selected in one of the high charts files, used to track down 
saturated stars and other artifacts passing as point sources in our catalog.
This tile was very specific to our observations and analysis and isn't useful
enough to make it's own affiliated package, but it demonstrates the power 
\emph{Toyz} gives its users to generate custom interactive content.

\section{Future Work}\label{Future}

\emph{Toyz} is still in its infancy and a number of exciting improvements
are planned for the future. The biggest upgrade will be integration
with Jupyter. Both iPython and Jupyter have
similar APIs that allow users to run python code from a web browser
but currently the interface to Jupyter is the traditional notebook
format. We're in the process of designing a notebook extension that
will implement the \emph{Toyz} workspace interface in iPython, which will
be useful for a wider community of users outside astronomy as well
as making it easier for end-users to develop their own custom
tiles and tools without extensive knowledge of javascript.

On the data visualization front there are plans to fully integrate \emph{Toyz}
with \emph{GLUE} to give users access to the entire \emph{GLUE} API in
a web browser, allowing them to connect to both local and remote data. We have
also been in contact with the \emph{Ginga} collaboration to discuss integrating
the current \emph{Toyz} ``slippy map'' viewer with the extensive toolset
developed by \emph{Ginga} to create a more complete image viewing platform.

Due to limitations in browser memory \emph{Highcharts} is limited in the
number of data points that can be displayed at once in an efficient
manner. More advanced technologies like WebGL are better suited
for the task of displaying large datasets as they work off of browser
plugins and expand the memory and functional capacities of web browsers.

As for \emph{Astro-Toyz}, we will continually be adding new features to incorporate
more \emph{astropy} functionality so that it can become a fully operational front-end
to the most useful \emph{astropy} functions, allowing undergraduates to process data in
classes with little to no programming background.

The source code and documentation for \emph{Toyz} is 
located at \url{https://github.com/fred3m/toyz} while \emph{Astro-Toyz}
can be downloaded from \url{https://github.com/fred3m/astro-toyz} where bug fixes or
and new pull requests are always welcome.

\section{Acknowledgements}\label{Acknowledgements}

We would like to thank the \emph{astropy} community for leading the way in
development of open source astronomical packages,
NSF grant AST-1313029 for supporting our research, and Cameron Bell for 
proofreading and providing helpful comments in the preparation of this
paper.

\section{References}

\bibliographystyle{elsarticle-harv.bst}
\bibliography{sources.bib}

\end{document}